\def\kms{$\rm km\;s^{-1}$}

\documentstyle[11pt,newpasp]{article}

\begin{document}

\title{Minor-Axis Stellar Velocity Gradients in Disk and Polar-Ring
Galaxies}
\author{F. Bertola \& E.M. Corsini}

\affil{Dipartimento di Astronomia, Universit\`a di Padova, Italy}

\section{Introduction}

Kinematical decoupling between two components of a galaxy suggests the
occurence of a second event.  In disk galaxy it is usually observed as a
counterrotation of the stellar disk with respect to the gaseous component
and/or with respect to a second stellar disk.  We discuss here three
cases of kinematical orthogonal decoupling of the innermost region of the
spheroidal component with respect to the disk or ring component in two Sa
spirals and in an elliptical galaxy with a polar ring.

\section{The Case Galaxies} 

\subsection{NGC~4698}

NGC~4698 is classified Sa by Sandage \& Tammann (1981) and Sab(s) by de
Vaucouleurs {\it et al.\/} (1991, RC3).  Sandage \& Bedke (1994, CAG) in The
Carnegie Atlas of Galaxies presented NGC~4698 in the Sa section as an example
of the early-to-intermediate Sa type.  They describe the galaxy as
characterized by a large central E-like bulge in which there is no evidence
of recent star formation or spiral structure.  The spiral arms are tightly
wound and become prominent only in the outer parts of the disk.

As discussed by Bertola {\it et al.\/} (1999), this galaxy is characterized
by a remarkable geometric decoupling between bulge and disk, whose apparent
major axes appear oriented in an orthogonal way at a simple visual inspection
of the galaxy images ({\it e.g.\/}, Panels 78, 79 and 87 in CAG).  The
$R-$band isophotal map of the Sa galaxy NGC~4698 shows that the inner region
of the bulge structure is elongated perpendicularly to the major axis of the
disk (Fig.~1), while the outer structure is perpendicular to the disk if a
parametric bulge-disk decomposition is adopted.  The bulge tends to become
rounder, but never elongated along the disk major axis, using a
non-parametric decomposition.

\begin{figure*}[ht] 
\includegraphics{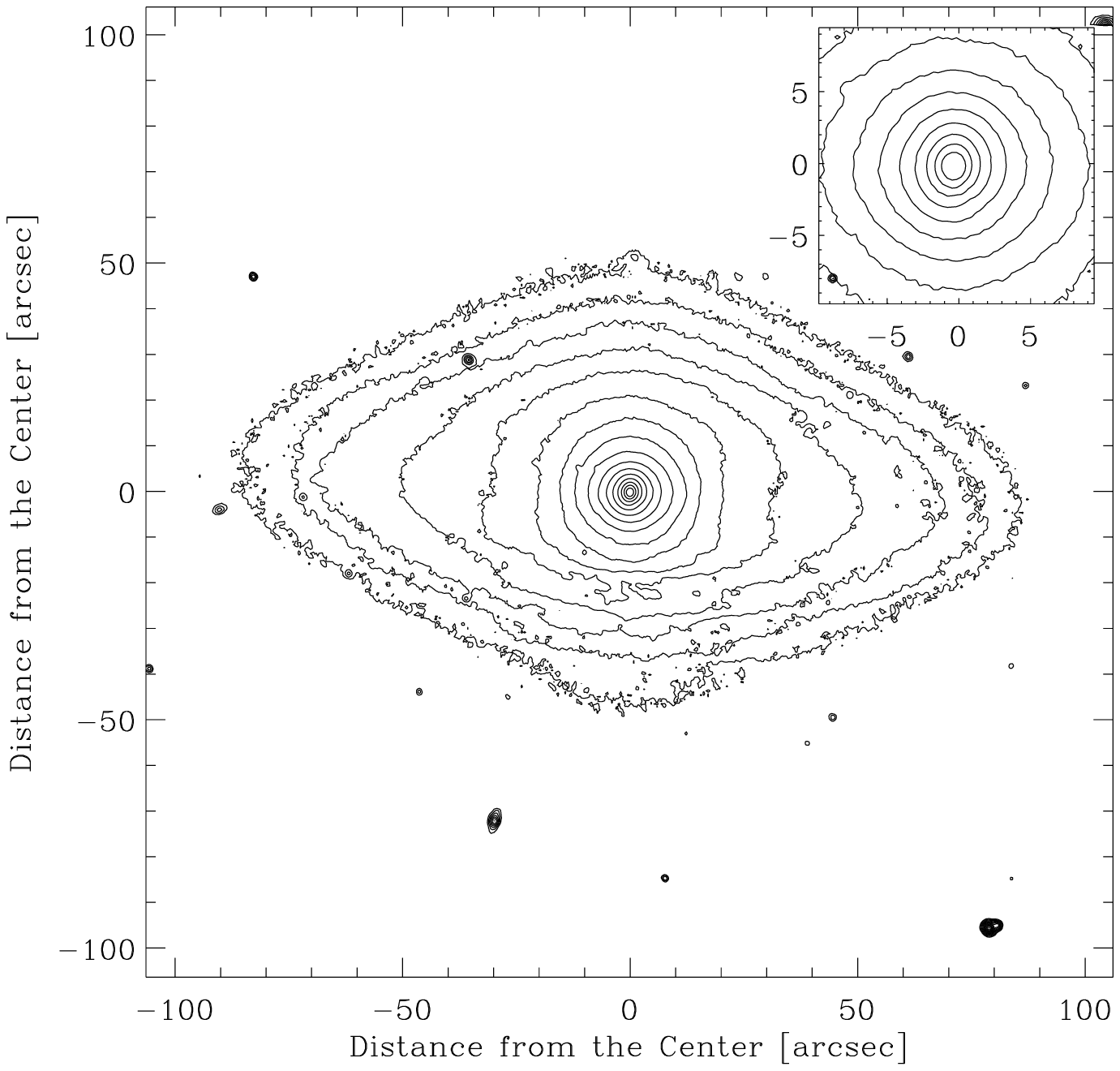}
\includegraphics{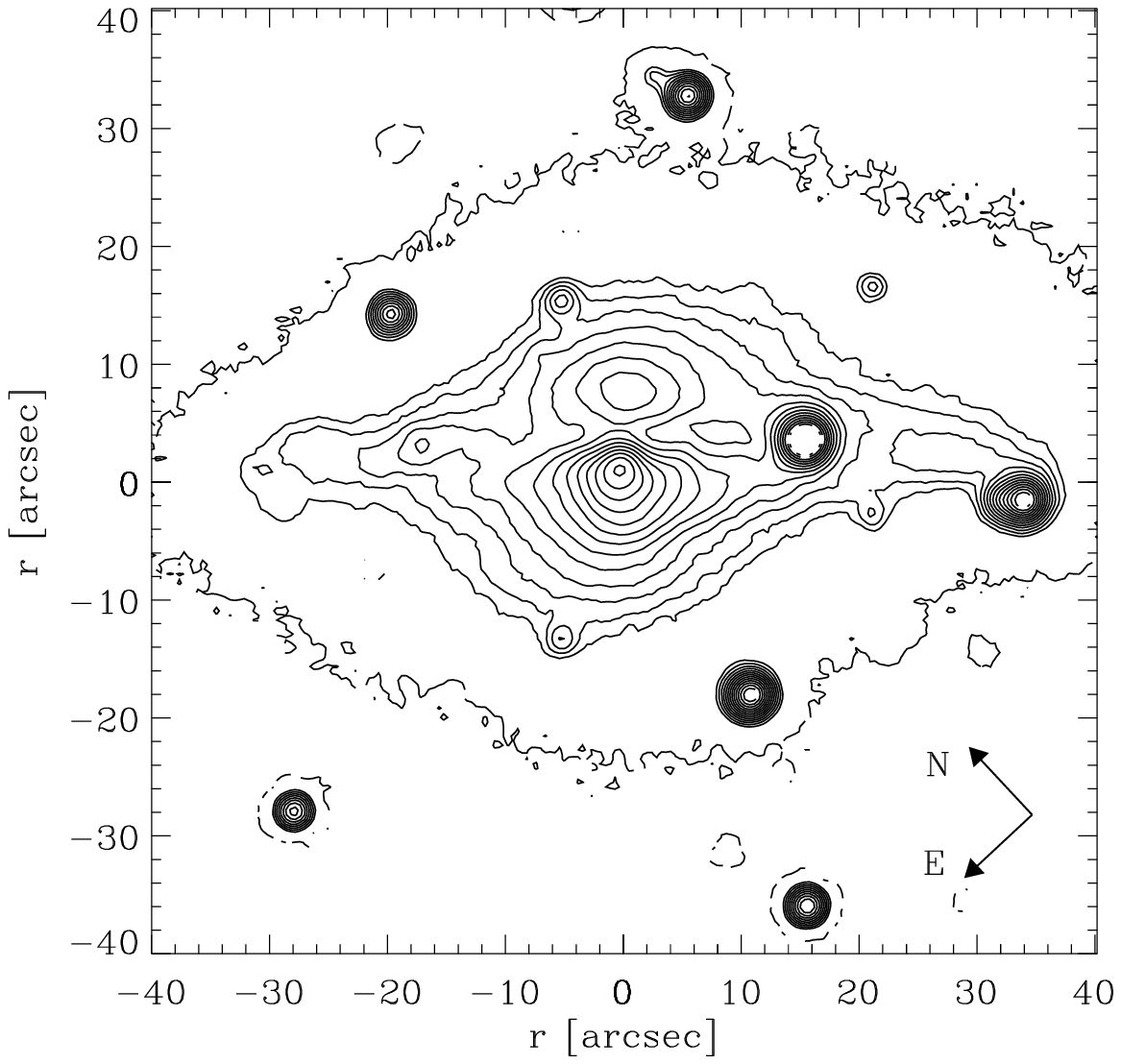}
\vspace{7cm} 
\caption{$R-$band isophotal map of NGC~4698 ({\it left panel\/})
and NGC~4672 ({\it right panel\/}).}
\end{figure*}

The stellar velocity curve measured along the major axis of NGC~4698 is
characterized by a central plateau, indeed the stars have a zero rotation for
$|r|\leq8''$ (Fig.~2).  At larger radii the observed stellar rotation
increases from zero to an approximately constant value of about 200 \kms\ for
$|r|\ga50''$ up to the farthest observed radius at about $80''$.  We measured
the minor-axis stellar kinematics out to about $20''$ on both sides of the
galaxy.  In the nucleus the stellar velocity rotation increases to about 30
\kms\ at $|r|\simeq2''$.  Then it decreases between $2''$ and $6''$ and it is
characterized by an almost zero value beyond $6''$.  The ionized-gas velocity
field, which is presented here for the first time (Fig.~2), is characterized
by a velocity gradient along the major axis higher than that of the stars.
Along the minor axis the gas velocities closely match those of the stars.
This suggests that this gas is associated to the stars giving rise to the
minor-axis velocity gradient.  These observations point out to the presence
in NGC~4698 of two gaseous and stellar components characterized by an
orthogonal geometrical and kinematical decoupling.

\begin{figure*}[ht] 
\includegraphics{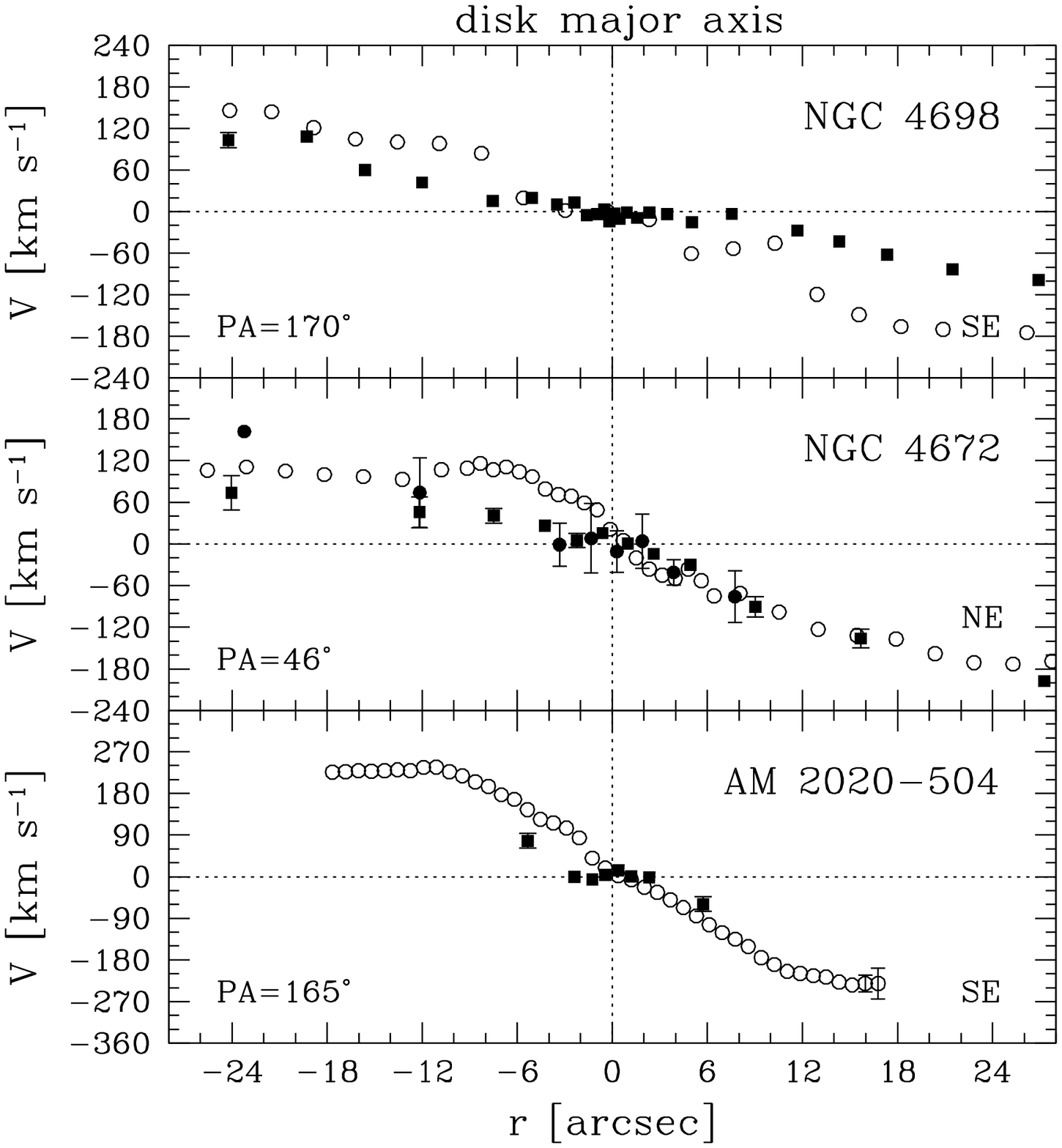}
\includegraphics{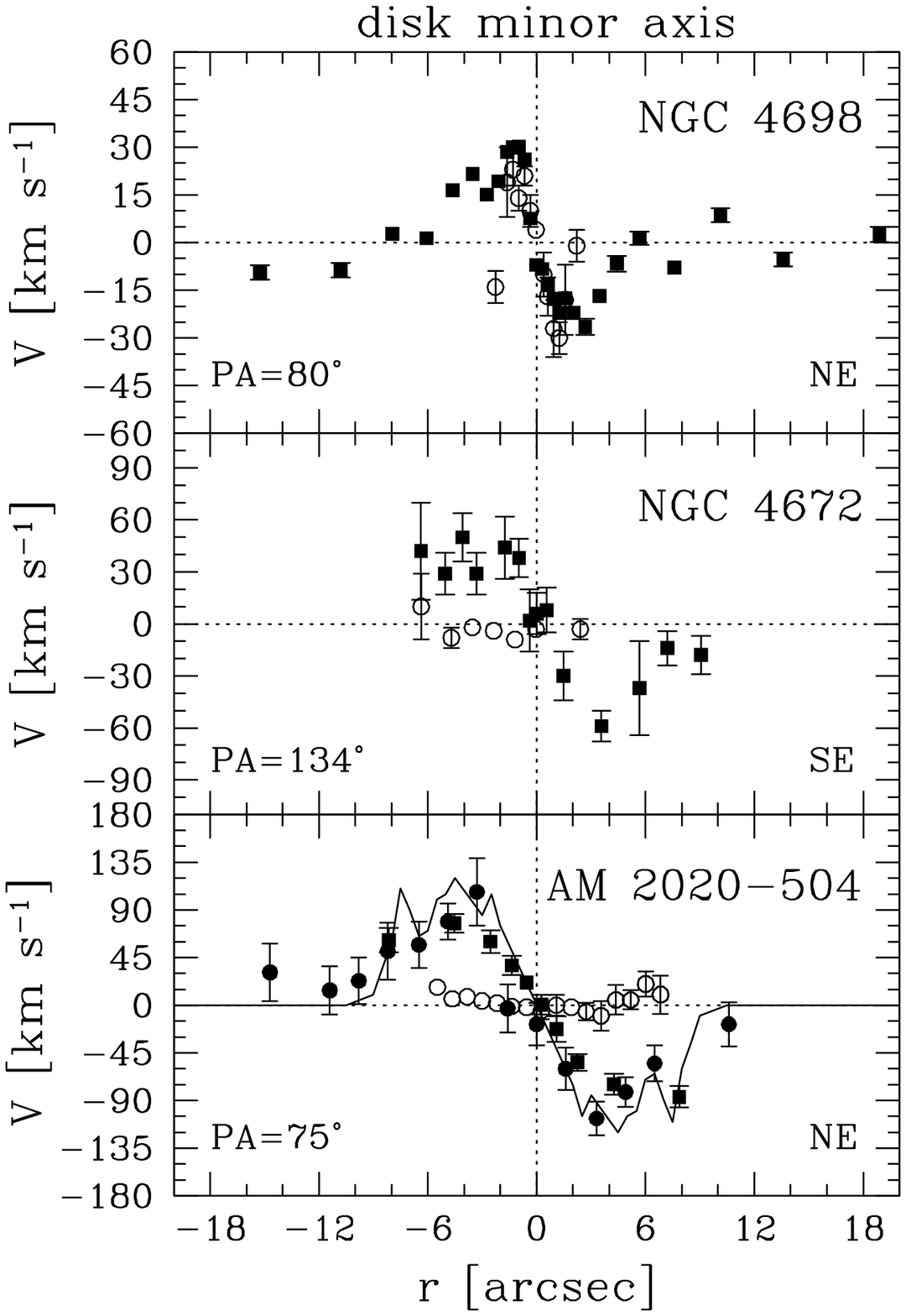}
\vspace{8cm} 
\caption{Stellar ({\it filled symbols\/}) and ionized-gas ({\it open
circles\/}) kinematics along the major ({\it left panel\/}) and minor ({\it
right panel\/}) axis of the disk (or ring) of NGC~4698, NGC~4672 and
AM~2020-504.  In the plot $1''$ corresponds to 56, 193, and 325 pc for
NGC~4698, NGC~4672 and AM~2020-504, respectively (assuming $H_0=75$ \kms\
Mpc$^{-1}$).  The stellar kinematics of NGC~4698 is taken from Bertola {\it
et al.\/} (1999).  The observations of NGC~4672 are from Sarzi {\it et al.\/}
(1999, in preparation).  For AM~2020-504 the {\it open circles} and {\it
filled squares} are new data, the {\it filled circles\/} and the {\it solid
line\/} represent the stellar velocities measured by Whitmore et al.  (1990)
and Arnaboldi {\it et al.\/} (1993a), respectively.}
\end{figure*}

\subsection{NGC~4672}

NGC~4672 is an highly-inclined early-type disk galaxy classified Sa(s) pec sp
in the RC3.  It is characterized by an intricate dust pattern crossing the
bulge near its center.  As for NGC~4698, the bulge of NGC~4672 appears
elongated in an orthogonal way with respect to the disk as shown by its
$R-$band isophotal map (Fig.~1).  Whitmore {\it et al.\/} (1990) considered
NGC~4672 a possible candidate for an S0 galaxy with a polar ring.  However,
as discussed in more detail by Sarzi {\it et al.\/} (1999, in this volume),
there are a number of evidences indicating that NGC~4672 is a spiral galaxy.

The major-axis stellar velocity curve is characterized by a central plateau
of zero rotation (Fig.~2).  The minor-axis stellar velocity curve shows a
steep gradient in the nucleus ($|r|\leq2''$), rising to maximum of about 80
\kms .  At larger radii the velocity tends to drop to a zero value.  The
major-axis ionized gas velocity curve (extending to about $60''$ from the
center on both sides of the galaxy) is radially asymmetric.  No significant
gas rotation is detected along the minor axis for $|r|<6''$.

Also in this case a kinematical orthogonal decoupling between the inner
stellar component and the disk is present.

\subsection{AM~2020-504}

This galaxy is considered the prototype of ellipticals with polar ring.  It
is constituted by two distinct structures:  a mostly gaseous ring and a
spheroidal stellar component (E4) with the major axes perpendicular each
other.  It has been extensively studied by Whitmore {\it et al.\/} (1990) and
by Arnaboldi {\it et al.\/} (1993a,b).

The most characteristic feature of the stellar kinematics of AM~2020-504 is
the presence of a velocity gradient along the major axis of the spheroidal
component and consequently perpendicular to the ring major axis.  This
gradient has been observed by Whitmore {\it et al.\/} (1990) and Arnaboldi
{\it et al.\/} (1993a) and confirmed by us (Fig.~2).  The high-resolution
spectrum by Arnaboldi {\it et al.\/} (1993a) shows a rise of the velocity up
to about $130$ \kms\ at $r\simeq4''$ followed by a decline to zero velocity
outside $10''$.  This, together with the zero velocity we observed within
$|r|<3''$ along the ring minor axis, indicates a rotation around the minor
axis.  Our velocity curve of the gas along the disk major axis is in
agreement with that of Whitmore {\it et al.\/} (1990).  The warped model for
the gaseous component discussed by Arnaboldi {\it et al.\/} (1993a) predicts
a velocity gradient along the minor axis of the ring, which is not shown by
our data.

\section{Discussion}

In spite of their morphological differences, the three galaxies described in
the previous section share two common characteristics:

\begin{itemize}

\item[{\it (i)\/}] The major axis of the disk (or ring) component forms an
angle of $90^\circ$ with the major axis of the bulge (or elliptical)
component.  This orthogonal geometrical decoupling is quite uncommon among
spiral galaxies.

\item[{\it (ii)\/}] The stellar kinematics along the disk minor axis
indicates the presence of a kinematically isolated core, which is rotating
perpendicularly with respect to the disk (or ring) component.

\end{itemize}

At this point we ask ourselves whether these galaxies also share similar
formation processes.  Arnaboldi {\it et al.\/} (1993a; see also Sparke 1986)
suggest that in AM~2020-504 the material forming the ring has been accreted
into polar orbits by an oblate elliptical.  The decoupled core may represent
material which has settled down in the simmetry plane of the oblate
spheroidal galaxy at the beginning of the acquisition process, and
subsequently has turned into stars.

If this mechanism produced also the kinematically isolated stellar cores in
the two spirals NGC~4698 and NGC~4672, then the geometrical orthogonal
decoupling observed in these galaxies results from the fact that the disk
moves in polar orbits around the central oblate spheroid.  No velocity
gradient along the disk minor axis and no geometrical decoupling are
expected if the acquisition process produces a disk settled on the
equatorial plane of the spheroidal component.  The rarity of bulges which
are perpendicularly sticking out from their disks would suggest that the
case of equatorial acquisition is the most common one.

The above considerations lead us to face a scenario in which the disk of a
spiral galaxy might have been formed as a second event by accretion around a
pre-existing bare spheroid.


\begin{references} 

\reference Arnaboldi, M., Capaccioli, M., Cappellaro, E., 
   {\it et al.\/} 1993a, A\&A, 267, 21 

\reference Arnaboldi, M., Capaccioli, M., Barbaro, G., 
   Buson, L.M. \& Longo, G. 
   1993b, A\&A 268, 103 

\reference Bertola, F., Corsini, E.M., Vega Beltr\'an, J.C., 
   {\it et al.\/} 1999, ApJ, 519, L127 

\reference Sandage, A. \& Bedke, J. 1994,
    The Carnagie Atlas of Galaxies (Washingthon: 
    Carnagie Institution, Flintridge Foundation) (CAG)
    
\reference Sandage, A. \& Tammann, G.A. 1981,
    A Revised Shapley-Ames Catalog of Bright Galaxies 
    (Washington: Carnegie Institution)

\reference Sparke, L. 1986, MNRAS, 219, 657

\reference Whitmore, B.C., Lucas, R.A., McElroy, D.B., 
   {\it et al.\/} 1990, AJ, 100, 1489 



\end{references}
\end{document}